\documentclass{PoS}
\newcommand{\be}{\begin{equation}}
\newcommand{\ee}{\end{equation}}
\newcommand{\beq}{\begin{eqnarray}}
\newcommand{\eeq}{\end{eqnarray}}
\newcommand{\bea}{\begin{eqnarray*}}
\newcommand{\eea}{\end{eqnarray*}}

\usepackage{epsfig}
\usepackage{epsf}
 \usepackage{dsfont}
\usepackage{slashed}
\usepackage{booktabs}
\usepackage{graphicx}
\usepackage{float}

\title{Exact calculation of disconnected loops}

\ShortTitle{Exact calculation of disconnected loops}

\author{C. Alexandrou\\
Department of Physics,  University of Cyprus, P.O. Box 20537, 1678 Nicosia,
Cyprus and Computation-based Science and Technology Research Center,
Cyprus Institute, P.O. Box 27456, 1645 Nicosia, Cyprus\\
        E-mail: \email{alexand@cyi.ac.cy}}
\author{D. Christaras\\
Department of Physics,  University of Cyprus, P.O. Box 20537, 1678 Nicosia,
Cyprus\\
        E-mail: \email{christaras.dimitrios@ucy.ac.cy}}
\author{\speaker{A. \'O Cais}\\
 Computation-based Science and Technology Research Center,
Cyprus Institute, P.O. Box 27456, 1645 Nicosia, Cyprus\\
        E-mail: \email{a.ocais@cyi.ac.cy}}
\author{A. Strelchenko\\
 Computation-based Science and Technology Research Center,
Cyprus Institute, P.O. Box 27456, 1645 Nicosia, Cyprus\\
        E-mail: \email{a.strelchenko@cyi.ac.cy}}

\abstract{
We present an implementation of the disconnected diagram contributions 
to quantities such as the flavor-singlet pseudoscalar meson mass which are accelerated by GPGPU technology utilizing  the NVIDIA CUDA platform. To enable the exact evaluation
of the disconnected loops we use a $16^3 \times 32$ lattice and $N_f=2$ Wilson
fermions simulated by the SESAM Collaboration. The disconnected loops 
are also computed using stochastic methods  with several noise reduction techniques.
 In particular, we analyze various dilution schemes as well as the recently proposed truncated solver method. We find consistency among the different methods
used for the  determination of the $\eta^\prime$ mass, albeit that
the gauge noise for the ensemble studied is large.  We also find that the effect of 'dilution' does not go beyond that of optimal statistical noise in many cases. It has been observed, however, that spin dilution does have a significant effect for some quantities studied. }

\FullConference{The XXVIII International Symposium on Lattice Field Theory, Lattice2010\\
		June 14-19, 2010\\
		Villasimius, Italy}

\begin{document}

\section{Introduction}

An accurate estimate of disconnected contributions to flavor singlet quantities remains one of the most computationally demanding problems
in hadronic physics.  
The most commonly adopted approach is to apply stochastic methods in order to estimate the quark propagator. A number of 
methods to reduce the stochastic noise inherent in such an approach has been developed and their respective merits investigated in detail in Ref.~\cite{Bali:2009hu}.
Such methods typically require large numbers of Dirac matrix inversions and hardware accelerators, 
such as graphics processors (GPUs), can dramatically accelerate these inversions~\cite{Barros:2008rd}.

The main goal of the present study is two-fold: Firstly,
 we compute the disconnected 
contribution to the flavor-singlet pseudo-scalar meson, $\eta^\prime$, mass which is also related to the  $U_A(1)$ anomaly in QCD.
Here, this is used as a case-study for the purposes of evaluating the efficacy of the implementation. 
Secondly, we examine the efficiency of various stochastic noise 
reduction techniques. More precisely, at this stage,
 we consider two techniques of variance reduction: partitioning (or dilution) \cite{Foley:2005ac} and the truncated solver method~\cite{Collins:2007mh}. 
We performed an \emph{exact} evaluation
 of the disconnected loops for $N_f=2$  Wilson 
fermions  on a lattice of size    $16^3\times 32$ using GPUs. 
The calculation is then
 repeated using stochastic methods. 
The exact calculation gives us an accurate benchmark by which to compare
 all stochastic variance reduction methods and explicitly exposes the gauge noise underlying each quantity to be measured.

\section{Lattice ensemble and simulation parameters}
For this exploratory study we use  $N_f = 2$  Wilson fermions
at  $\beta = 5.6$  and hopping parameter $\kappa = 0.157$, which
corresponds to pion mass of $m_\pi = 884$~MeV on
a lattice of size $16^3 \times 32$~\cite{SESAM}. The   
 lattice spacing is $a = 0.08$~fm as determined from the nucleon mass
at the physical point~\cite{Alexandrou}.
For constructing the meson propagators we utilized both local and smeared 
quark fields. In the latter case, we apply gauge-covariant Gaussian 
smearing using a range of smearing parameters. 
 
The stochastic estimate of the disconnected  quark loops is performed  
using complex  $Z_2$ noise for the source vectors 
in combination  with several partitioning (dilution) schemes and the truncated
solver method~\cite{Bali:2009hu}.   Specifically, 
 we consider various combinations of space, spin and color dilution schemes. Colour dilution leads to a multiplicative factor of 3 for the number of inversions. In spin space, a full dilution leads to a factor of 4 for the inversions. In this case an even-odd partitioning of the space can alternatively be employed leading to an increase of a factor of 2 in the number of inversions. For spatial dilutions, in addition to an even-odd dilution, we have also applied a \emph{cubic} dilution, where separate sources are placed on each vertex of an elementary 3-d cube and repeated throughout the lattice, leading to an increase of a factor of 8 in the number of inversions. Time dilution
is applied in all cases and translational invariance exploited so that this does not increase the number of required inversions.

The truncated solver method~\cite{Bali:2009hu} effectively partitions the problem into a low precision and high precision space. A large number
of low precision inversions are carried out to achieve an approximation to the propagator with low stochastic error (but only accurate to low precision). A high precision stochastic correction is then applied using a small ensemble with the corresponding inversions carried out to high precision.  We use a stochastic ensemble of 5000 noise vectors for the low precision space with an ensemble of 500 noise vectors for the high-precision correction. The inversion tolerance 
for the low precision was chosen to be $10^{-6}$ such that one 
can restrict oneself to  a single precision conjugate gradient inversion (which is very efficient on GPU accelerators), 
while in the case of full precision 
the tolerance was set to $10^{-10}$. The ensemble sizes were chosen to be quite large in order to avoid any quantity-specific tuning of the ensemble sizes. 

Finally, as was already mentioned,  the
exact evaluation of the all-to-all propagator is also carried out. This is clearly the most computational intensive part, and was only possible due to the use of graphics accelerators employing the QUDA library (as was used for all inversions), which
  provides mixed precision implementations of CG and BiCGstab solvers for 
the NVIDIA CUDA platform~\cite{Clark:2009wm}. This provides a benchmark at the level of gauge-noise for all quantities with contributions from disconnected loops.

\section{Results}

For all-to-all propagators,  a general isovector two-point correlation function, $C^{BA}({\bf p}, \Delta t)$, for the creation of a particle at timeslice $t$ with momentum $p$
from the operator $\Gamma_A$ and its annihilation at timeslice $t+\Delta t$ with the operator $\Gamma_B$ is given by,
\begin{equation}\label{eqn:all_corr}
C^{BA}({\bf p}, \Delta t) = - \frac{1}{L^3T}\sum_{{\bf x,y,} t}\langle \textrm{Tr}(S_{F}({\bf y},t;{\bf x},t+ \Delta t)\Gamma^B({\bf p}) S_{F}({\bf x},t+ \Delta t;{\bf y},t)\Gamma^A
({\bf p})) \rangle,
\end{equation}
where $S_{F}({\bf x},t;{\bf x^\prime},t^\prime)$ is the propagator from spacetime point $({\bf x},t)$ to spacetime point $({\bf x^\prime},t^\prime)$, spin and colour indices are suppressed and phases for momentum projections (and quark smearing operations) are incorporated into the definition of the operators $\Gamma_A$ and $\Gamma_B$. For isoscalar quantities, disconnected loops give a contribution $D^{BA}({\bf p}, \Delta t)$ to the correlation function,
\begin{equation}\label{eqn:all_dis_corr}
D^{BA}({\bf p}, \Delta t) = - \frac{1}{L^3T}\sum_{{\bf x,y,} t}\langle \textrm{Tr}(S_{F}({\bf x},t;{\bf x},t)\Gamma^B({\bf p}) S_{F}({\bf y},t+ \Delta t;{\bf y},t+ \Delta t)\Gamma^A
({\bf p})) \rangle,
\end{equation}
In our particular case of the $\eta^\prime$ meson in an $N_f=2$ gauge ensemble, if we suppress all operator and momentum indices, 
\begin{equation}
C_{\eta^\prime}(t) = C_{\pi}(t) - 2D(t).
\end{equation}

For mesons on lattices with periodic boundary conditions $C(t) \sim  e^{-mt}+e^{-m(N_T-t)} $ for $t$ large (where $N_T$ is the lattice temporal extent). We can therefore analyze the ratio of the disconnected quark loop, $D(t)$, 
and connected correlation function, $C_{\pi}(t)$, to extract the flavour-singlet pseudoscalar meson mass,
\begin{equation}
\frac{D(t)}{C_{\pi}(t)} \sim_{t\rightarrow \infty} A - B\frac{e^{-m_{\eta^\prime} t} + e^{-m_{\eta^\prime} (N_T-t)}}{e^{-m_\pi t} + e^{-m_\pi (N_T-t)}},
\end{equation}
where $m_\pi$ and $m_{\eta^\prime}$ are the masses of the $\pi$ and $\eta^\prime$ mesons and $A$, $B$ are additional fit parameters.  
$m_\pi$ can be determined separately to the $1\%$ level and inserted as a prior leaving only 3 parameters in the fit function. 
This approach also accounts for independent smearing of the connected and disconnected loops.

\begin{figure}[h!]
\begin{center}
{\includegraphics[angle=-90, width=0.8\linewidth]{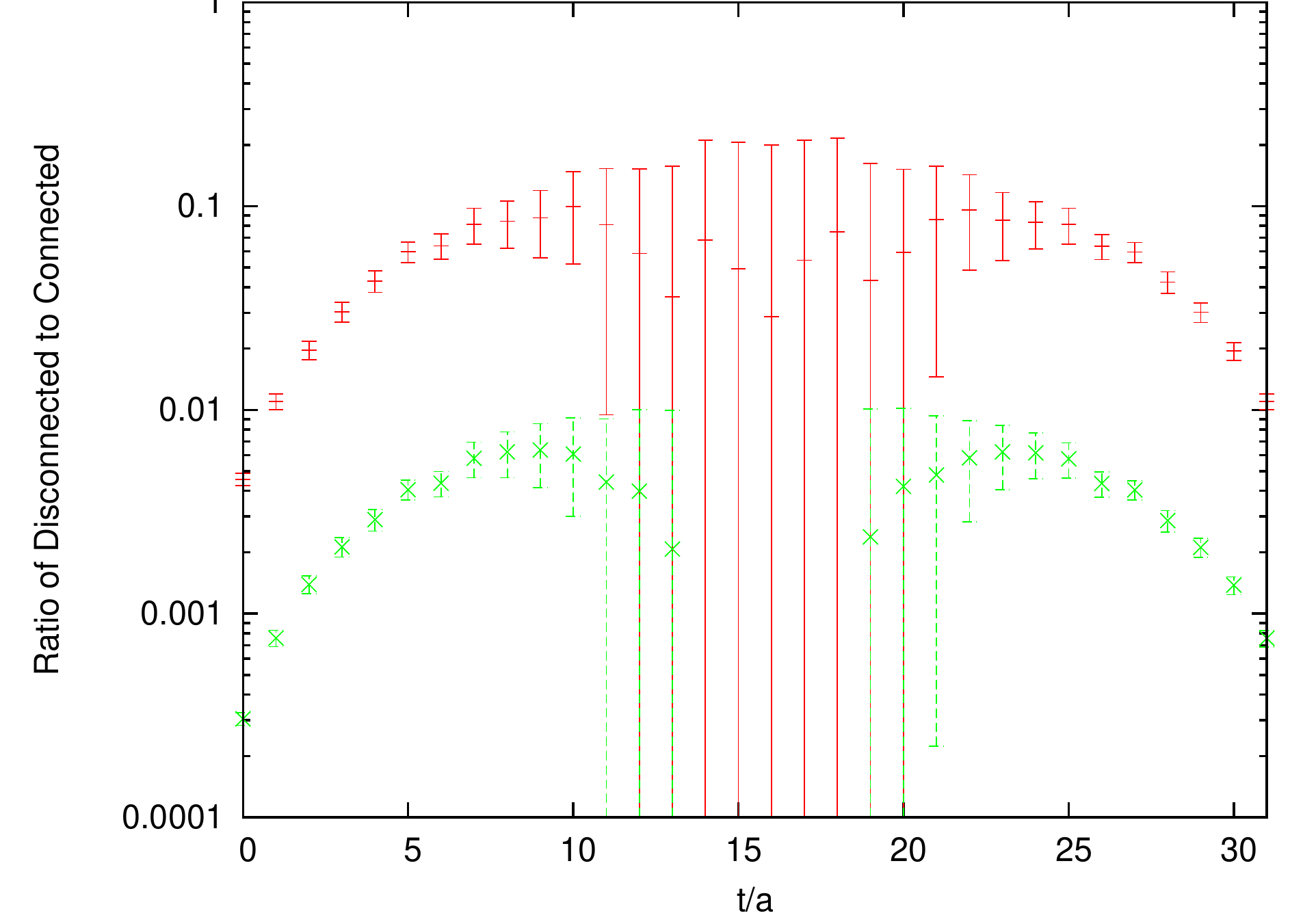}}
\end{center}
\caption{Ratio $\frac{D(t)}{C_{\pi}(t)}$ for the $\eta^\prime$ meson computed using exact approach (without smearing in red and with smearing in green colours).}
\label{fig:ratio1}
\end{figure}
In the case of the exact evaluation, the only source of error
 comes from the statistical error of the gauge ensemble, and
therefore we will employ this fact to assess the results obtained using 
stochastic methods. We compare exact results for the ratio $D(t)/C_{\pi}(t)$ obtained with local quark field operators
 with those obtained with smeared quark fields in
Fig.~\ref{fig:ratio1}. Given that $m_\pi$ can be determined to the $1\%$ level it is clear that the gauge noise derived from the disconnected loops in this quantity is large.
A naive fitting of the data constrained between $t_{min} = 2a$ and $t_{max}=11a$  gives 
a value for the mass $am^L_{\eta^\prime} = 0.41 \pm 0.04$ and $am^S_{\eta^\prime} = 0.40\pm0.05$ for local  
and smeared operators, respectively. Here we used $am_{\pi} = 0.3454(19)$. 
On the other hand, fitting in the range $t_{min} = 3a$ and $t_{max}=11a$  
 gives us accordingly $am^L_{\eta^\prime} = 0.51\pm0.07$ and $am^S_{\eta^\prime} =  0.49\pm0.09$.  

\begin{figure}[h!]
\begin{center}
{\includegraphics[angle=-90, width=0.8\linewidth]{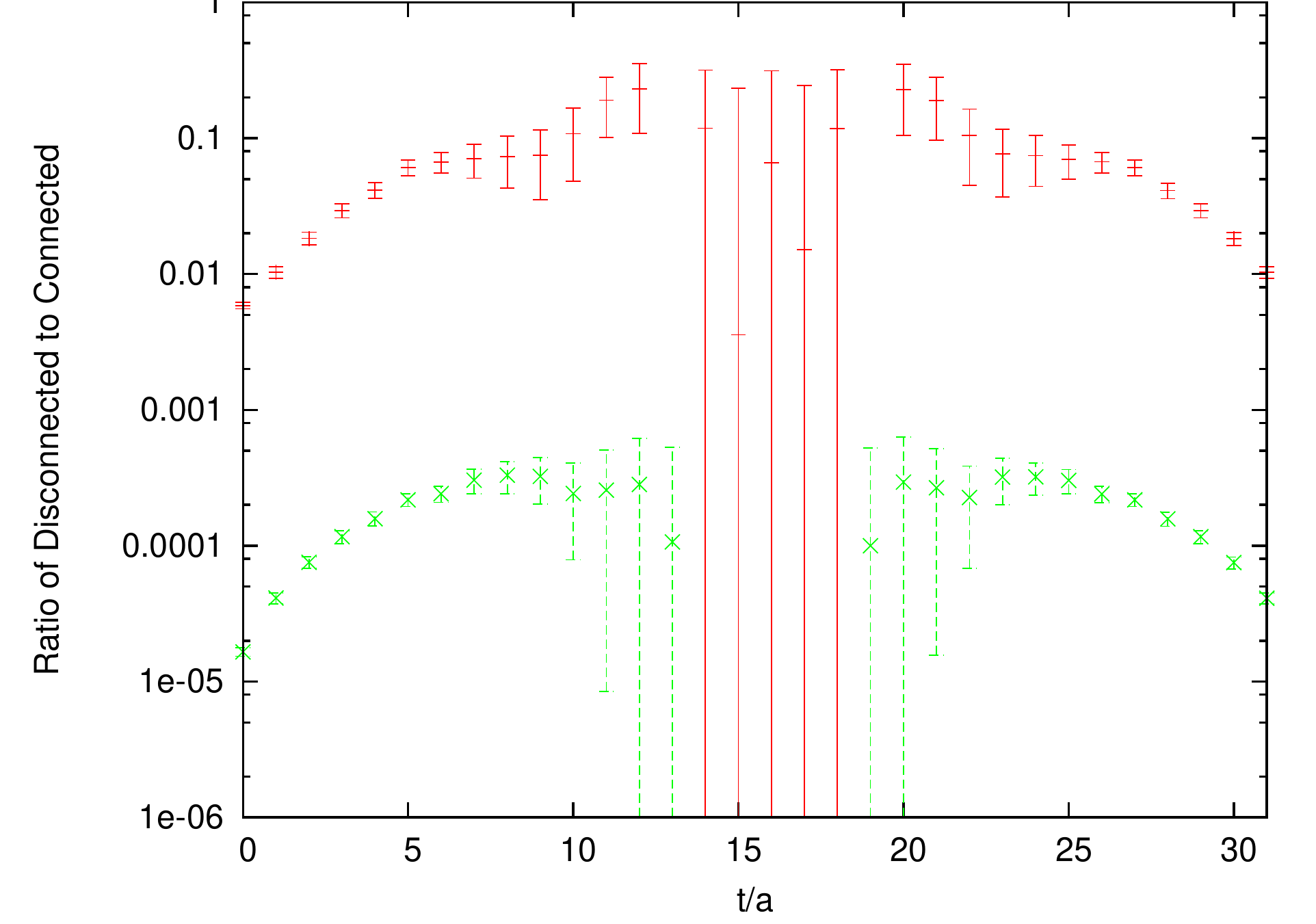}}
\end{center}
\caption{Ratio $\frac{D(t)}{C_{\pi}(t)}$ for the $\eta^\prime$ meson computed using TSM (without smearing in red and with smearing in green colours).}
\label{fig:ratio2}
\end{figure}
If we look at the same quantities using the truncated solver method, the corresponding plots in logarithmic scale 
(for local and smeared operators) are given in Fig.~\ref{fig:ratio2}. 
The results of fitting in the range $t_{min} = 2a$ and $t_{max}=10a$ gives
$am^L_{\eta^\prime} = 0.42\pm0.07$ for local fields and $am^S_{\eta^\prime} = 0.42\pm0.06$ 
for smeared fields, respectively. Fitting in the range $t_{min} = 3a$ and $t_{max}=10a$ 
 provides the following estimates for the pseudo-scalar mass:  $am^L_{\eta^\prime} = 0.57\pm0.11$ and $am^S_{\eta^\prime} = 0.51\pm0.1$. 
These results are summarized in Table~\ref{tab:results}. The value of $am_{\eta^\prime}=0.40(5)$ is relatively consistent
with the results obtained using $N_f=2$ twisted mass fermions at $m_\pi\sim 500$~MeV~\cite{Jansen:2008wv}.
\begin{table}
\label{tab:results}
\caption{The $\eta^\prime$ mass using exact and the truncated solver method (TSM)  for the evaluation of the disconnected loop.}
\begin{center}
{\small \vskip 0.5cm \noindent\begin{tabular}{|l|c|c|c|c|} \hline & $t_{min}$ & $t_{max}$ & $am^L_{\eta^\prime}$ & $am^S_{\eta^\prime}$  \\\hline 
Exact & 2 & 11 & $0.41 \pm 0.04$ & $0.40\pm0.05$
\\\hline 
Exact & 3 & 11 & $0.51\pm0.07$ & $0.49\pm0.09$\\\hline
TSM & 2 & 10 & $0.42\pm0.07$ & $0.42\pm0.06$
\\\hline 
TSM & 3 & 10 & $0.57\pm0.11$ & $0.51\pm0.1$\\\hline 
\end{tabular}
}
\end{center}
\end{table}

\begin{figure}[h]
\begin{minipage}{0.58\linewidth}
\hspace{-0.2\linewidth}
{\includegraphics[angle=-90,width=\linewidth]{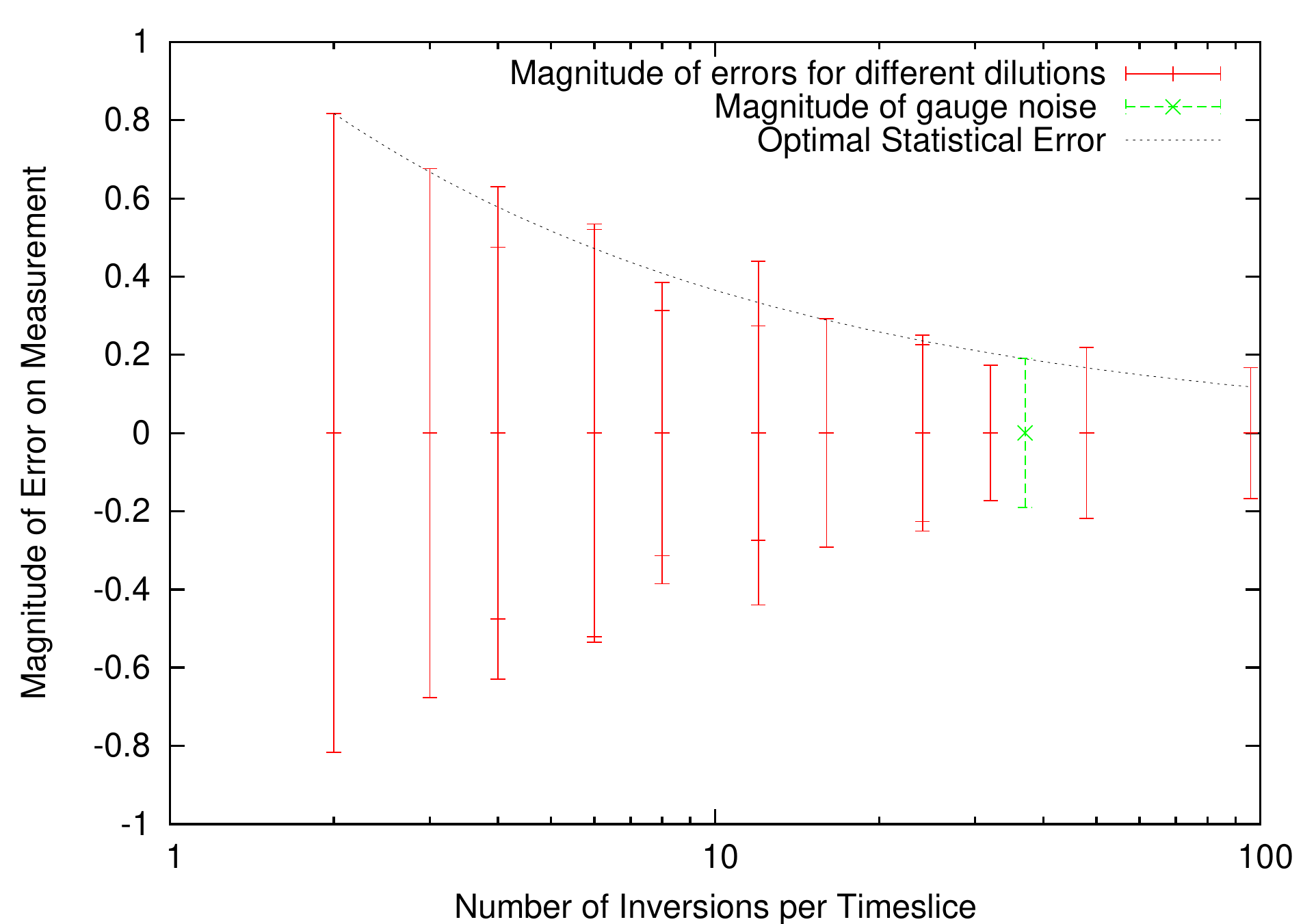}}
\end{minipage}
\begin{minipage}{0.58\linewidth}
\hspace{-0.17\linewidth}
{\includegraphics[angle=-90,width=\linewidth]{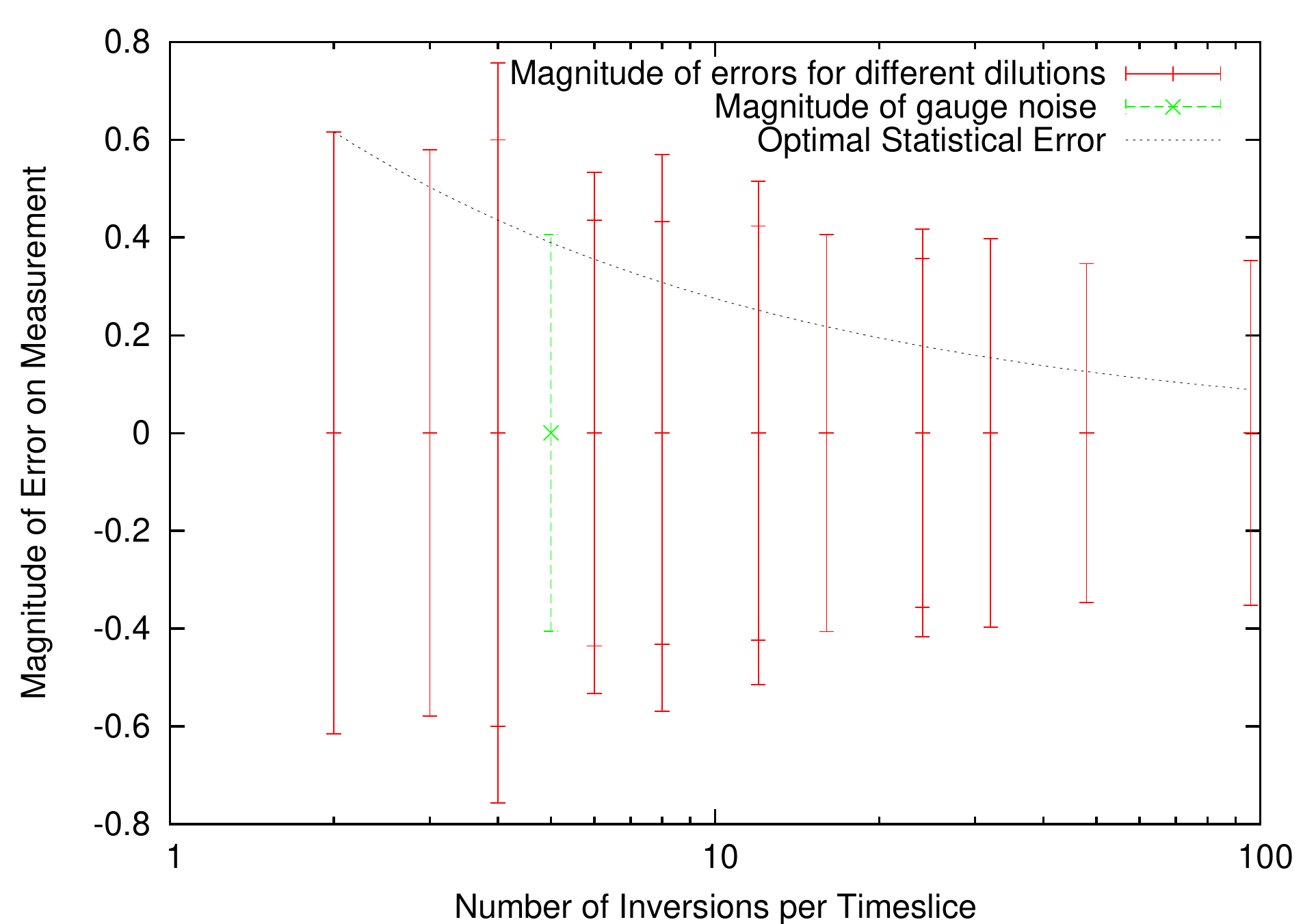}}
\end{minipage}
\caption{Magnitude of errors for $Tr(S_F(x, x) \Gamma)$. Right: for the $\gamma_5$-operator (gauge level noise achievable with 37 inversions); Left: for the identity operator (gauge level noise achievable with 5 inversions).}
\label{fig:errors}
\end{figure}
We have also considered the stochastic estimate of the disconnected diagrams
 using  $Z_2$ noise and 2 different approaches. First, we examine the number of noise vectors
 required such that one can reach a level of stochastic accuracy consistent with the statistical noise of the gaugefield ensemble. 
To this end, we have inspected the trace of zero-momentum projected disconnected loop 
 $Tr(S_F(x, x) \Gamma)$, for a range of operators $\Gamma$.
For example, for the particular case of a dilution approach with a $\gamma_5$ operator
 insertion, the left of Fig.~\ref{fig:errors} shows the
 dependency of the magnitude of the error in the trace on the number of noise vectors. This figure shows the number of inversions, $N_{inv}$, required for each dilution scheme along the $x$-axis, e.g., full colour dilution with full spin dilution would require 12 inversions (a factor of 3 for each colour and 4 for each spin, as described earlier). As a reference, we also plot the optimal statistical error (behaving as $\frac{1}{\sqrt{N_{inv}}}$) extrapolating from the first data point to show the expected behaviour of increasing the ensemble size. Finally we insert the gauge-level noise (from the exact calculation) at the point where this gauge error is consistent with overall error from the optimal error extrapolation. As one can see, in the case of the $\gamma_5$-operator the dilution approach is consistent with the optimal statistical error and one needs at least $37$ inversions to achieve gauge noise accuracy. A similar analysis can be done for other disconnected loops, on the right of Fig.~\ref{fig:ratio1} we show results for an identity operator insertion,
 where the gauge noise can be reached with just $5$ noise vectors and, consequently, dilution can have no beneficial effect for the measurement. Clearly, the size of the stochastic ensemble required is operator-dependent, as noted in Ref.~\cite{Bali:2009hu}.  

\begin{figure}[h]
\begin{minipage}{0.58\linewidth}
\hspace{-0.2\linewidth}
{\includegraphics[angle=-90,width=\linewidth]{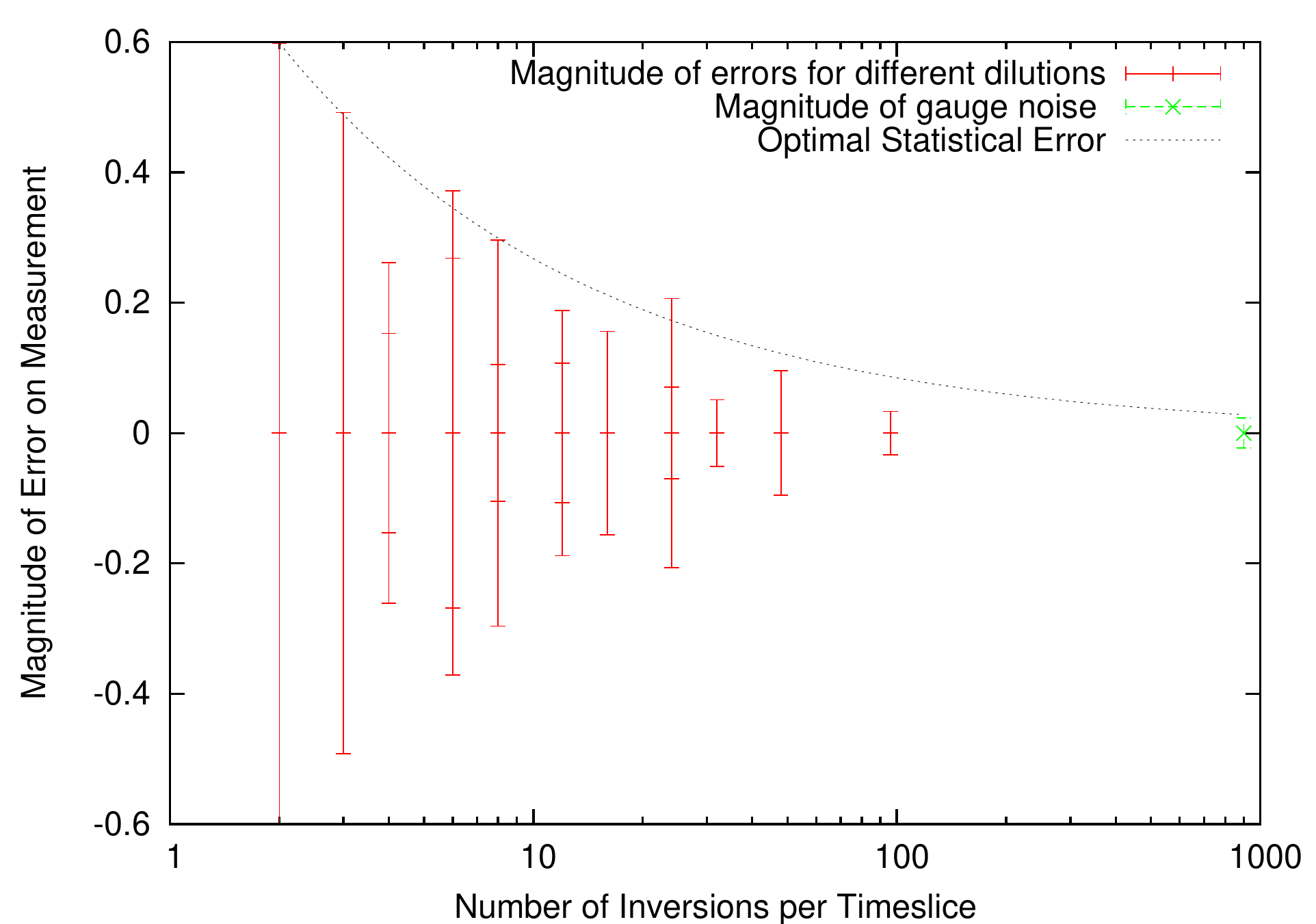}}
\end{minipage}
\begin{minipage}{0.58\linewidth}
\hspace{-0.17\linewidth}
{\includegraphics[angle=-90,width=\linewidth]{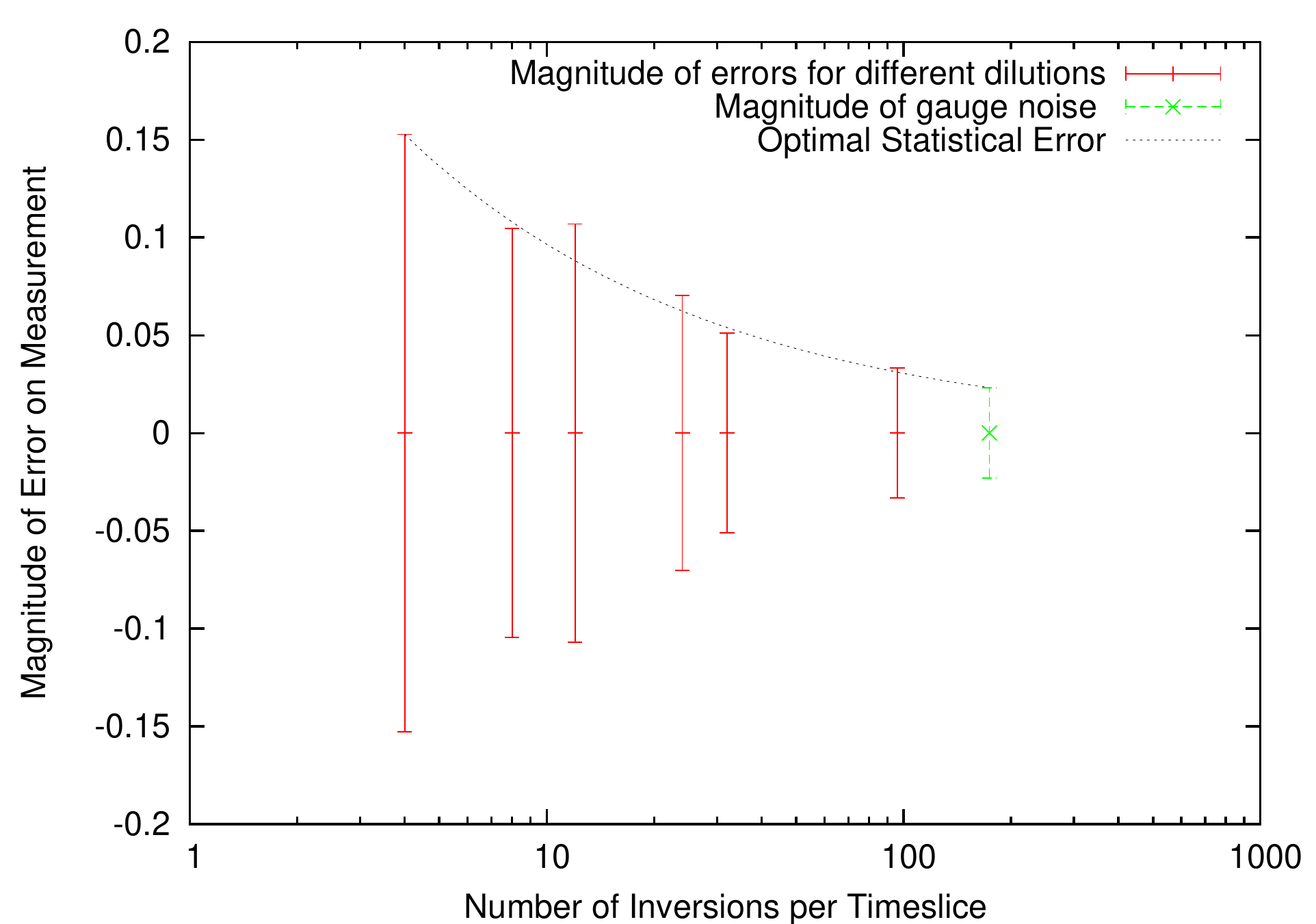}}
\end{minipage}
\caption{Magnitude of errors for $Tr(S_F(x, x) \gamma_1 \gamma_3)$. Left: for all dilution schemes; Right: for all dilution schemes which include full spin dilution (gauge level noise achievable with 174 inversions)}
\label{fig:spin}
\end{figure}

In Fig.~\ref{fig:spin}, we show similar plots for a $\gamma_1\gamma_3$ operator insertion. On the left we show an identical plot to those in Fig.~\ref{fig:errors} with the exception that the gauge noise cannot be reached within a sample size of 1000 and the gauge noise is simply plotted near this limit for illustrative purposes. On the right we plot the same data for the specific cases where full spin dilution is used. As can be seen, spin dilution has, in this case, a dramatic effect that allows the achievement of gauge level noise within 174 inversions. This effect is likely due to the strong off-diagonal nature of this gamma combination in this basis and has been also been observed in other quantities.       Also, we again observe that dilution behaves consistently with the optimal statistical error.

\section{Summary}

We have computed the disconnected contribution to $\eta^\prime$ meson mass 
using both exact and stochastic evaluation. 
 Stability in the fit region has not been observed and the level of noise from the gaugefield ensemble is large, particularly at large separations. We have also compared the truncated solver method against the exact approach in our attempt to evaluate the $\eta^\prime$ meson mass and find results consistent though non-conclusive between the two approaches. This is of course due to the gaugefield ensemble noise inherent in the quantity for the sample.

We analyzed efficiency of different noise reduction techniques using the gaugefield ensemble noise from the exact evaluation as a benchmark. We find that, on these lattices, for the stochastic methods the gauge noise becomes the dominant source of the error already for $37$ inversions in the case
of $Tr(S_F(x, x) \gamma_5)$ and just $5$ inversions in the case of $Tr(S_F(x, x))$. The statistical error from using the dilution approach appears to behave similarly to statistical noise in many cases for these types of operators. In particular cases however, such as $Tr(S_F(x, x)\gamma_1\gamma_3)$, spin dilution has been seen to have a significant effect.

\begin{acknowledgments}
Alan \'O Cais is supported by the Research Promotion Foundation of Cyprus under grant $\Pi$PO$\Sigma$E$\Lambda$KY$\Sigma$H/$\Pi$PONE 0308/09.
The production runs were carried out on an 8-node Tesla cluster at the Cyprus Institute, funded under the same grant. Additional production runs were carried out on the Lincoln cluster at the National Center for Supercomputing Applications at the University of Illinois. 
\end{acknowledgments}

\end{document}